\documentstyle[epsf]{ioplppt}                
\begin{document}
\jl{1}
\title{A new method to calculate the spin-glass order parameter
 of the two-dimensional $\pm J$ Ising
model$\sim $}[A new method to calculate the spin-glass order parameter]

\author{Hidetsugu Kitatani\ftnote{1}{E-mail address:kitatani@vos.nagaokaut.ac.jp}
 and Akira Sinada}

\address{Department of Electrical Engineering, Nagaoka University of
Technology, Nagaoka, Niigata 940-2188, Japan}

\begin{abstract}
A new method to numerically calculate  the $n$th moment of the
spin overlap  of 
the two-dimensional $\pm J$ Ising model is developed
using the identity derived by one of the authors (HK) several
years ago.
By using the method,
the $n$th moment of the spin overlap
can be calculated as a simple average of
the $n$th moment of the total spins
with a modified bond probability distribution.
The values of the Binder parameter etc have been extensively calculated  
with the linear size, $L$, up to $L=23$.
The accuracy of the calculations in the present method  is
similar  to that 
in
the conventional transfer matrix method with about $10^{5}$ bond samples.
The simple scaling plots of the Binder parameter and the 
spin-glass susceptibility indicate the existence of a
finite-temperature spin-glass phase transition.
We find, however, that
 the estimation of  $T_{\rm c}$
is strongly affected by the corrections to scaling within
the present data ($L\leq 23$).
Thus, there still remains the possibility that $T_{\rm c}=0$,
contrary to the recent results which suggest the existence
of a finite-temperature spin-glass phase transition.
\end{abstract}

%
%

\section{Introduction}
Over the last two decades, investigations of 
spin-glass problems
have been  extensively performed[1-20].  It is now widely believed
that the three-dimensional $\pm J$ Ising model shows a 
finite-temperature spin-glass phase transition [1-6], while the critical 
temperature
of the two-dimensional $\pm J$ Ising model
is zero [5-11].
Most of these studies have been done using  Monte Carlo simulations,
where  the thermal relaxation time
in the simulations becomes very large in the low-temperature
region. This makes 
the investigations of the two-dimensional models 
rather difficult, 
since the calculations of the physical
quantities have to be performed at very low temperature.
In previous studies, the data in the finite-size scaling analysis
were in good agreement with
a scaling function with the critical temperature, $T_{\rm c}=0$.
 The results, however,
have not completely excluded the possibility of 
a finite-temperature spin-glass phase transition.
Recently, Shirakura {\it et al}[12-15]
 have deduced the
existence of a finite-temperature spin-glass phase transition
of the two-dimensional models
using extensive Monte
Carlo simulations. 
To clarify the critical properties of the two-dimensional
$\pm J$ Ising model, more precise results in the low-temperature
region are necessary.

The transfer matrix method is free from 
the problem of thermal equilibration, and has been very
successfully used to determine the ferromagnetic-nonferromagnetic 
phase boundary of the two-dimensional $\pm J$ Ising model
in the $p-T$ plane 
 \cite{Ozeki,Kitatani3}
($p$ is the concentration of the ferromagnetic bond).
But,
for the problem of the spin-glass phase transition, 
the transfer matrix method 
has only been  used for the calculations of defect energies
and correlation functions
\cite{Morgenstern,Ozeki2},
and has not been widely used for direct calculations
of the $n$th moment of the spin ovelap
(the spin-glass susceptibility, the Binder parameter, etc)
since, when we use real replicas in  the calculations,
 the maximum linear size applicable
  is one-half of that 
 in the calculations of the $n$th moment of the total spins.  
Thus, so far,  no
extensive result for the spin-glass phase transition
 with 
direct calculation of the spin-glass susceptibility etc
has been given by the transfer matrix method.

In this paper, we present a new method to numerically calculate 
the $n$th moment of the spin overlap of
the two-dimensional $\pm J$ Ising model,
using the identity derived by one of the present authors
several years ago \cite{Kitatani}.
In this identity,
the $n$th moment of the spin overlap is transformed as a simple
average of the $n$th moment of the total spins
with a modified bond probability
distribution. 
Following a newly developed process, explained in section 2, 
 we successively make the bond configurations
according to the modified bond probability distribution
using the Monte Carlo technique.
In each bond configuration, we
calculate the $n$th moment of the total spins 
 by the transfer matrix method.

We have performed extensive 
calculations
of the $n$th moment of the spin overlap up to the linear size, $L=23$.
The accuracy of the calculations in the present method
is similar  to that in the conventional
transfer matrix method  with 
about $10^{5}$
bond samples. Thus, the statistical errors in
the present study are 
about an  order of magnitude smaller
than those in previous studies. Therefore, we can analyse
the obtained data in detail using 
finite-size scaling analysis including the corrections to scaling.
Our results show that the estimation of $T_{\rm c}$
is strongly affected by the corrections to scaling
in the two-dimensional $\pm J$ Ising model.
Thus, there still remains the possibility that $T_{\rm c}=0$,
contrary to the recent results by Shirakura {\it et al}
\cite{Shirakura,Shiomi,Matsubara} which suggest the existence
of a finite-temperature spin-glass phase transition.

\section{New calculation method for the spin-glass order parameter}
We consider the two-dimensional $\pm J$ Ising model on 
an $L\times L$ square lattice with only nearest neighbour interactions.
The Hamiltonian  is written as follows:
\begin{equation}
 {\cal H} = -\sum_{(ij)}\tau_{ij}S_{i}S_{j}, 
\end{equation}
where $S_{i}=\pm 1$, and the
summation of $(ij)$ runs over all the nearest neighbours.
We take the skew boundary condition in one direction, and the free
boundary condition in the other direction.
 Each $\tau_{ij}$ is determined according to the following
probability distribution:
\begin{equation}
    P(\tau_{ij}) = p\delta (\tau_{ij}-1)+(1-p)\delta (\tau_{ij}+1).
\end{equation}
In this paper, we make $J=1$.
We define the spin overlap, $Q$, between two replicas in each bond
configuration as
\begin{equation}
  Q = \sum_{i=1}^{N} S_{i}^{\alpha }S_{i}^{\beta },
\end{equation}
where $\alpha$ and $\beta$ denote the two replicas, and
$N$ is the total number of spins.
When we define $K_{p}$ as
\begin{equation}
     \exp (2K_{p}) = \frac {p}{1-p},
\end{equation}
the $2n$th moment of the spin overlap is written as
\begin{equation}
\fl    [<Q^{2n}>_{{T},\{S^{\alpha},S^{\beta}\}}]_{p}=
    \frac {1}{(2\cosh (K_{p}))^{N_{\rm B}}}
   \sum_{\tau_{ij}=\pm 1}\exp (K_{p}\sum_{(ij)}\tau_{ij})
   <Q^{2n}>_{{T},\{S^{\alpha},S^{\beta}\}},
\end{equation}
where $<\cdots >_{{T},\{S^{\alpha},S^{\beta}\}}$ denotes the thermal average
both for the $"S^{\alpha}"$ and $"S^{\beta}"$ spins
in a bond configuration, $\{\tau \}$ at temperature, $T$.
$[\cdots ]_{p}$ denotes the configurational average at the
ferromagnetic bond concentration, $p$,
and $N_{\rm B}$ is the number of bonds \cite{Nishimori}.
By the use of a  local gauge transformation,
an identity has been derived \cite{Kitatani}:
\begin{equation}
\fl     [<Q^{2n}>_{{T},\{S^{\alpha},S^{\beta}\}}]_{p}=
    \frac {1}{(2\cosh (K_{p}))^{N_{\rm B}}}
   \sum_{\tau_{ij}=\pm 1}\exp (K\sum_{(ij)}\tau_{ij})
    \frac {Z(K_{p},\{\tau \})}{Z(K,\{\tau \})}<M^{2n}>_{{T},\{S\}},
\end{equation}
where $M$ denotes the total spins,
\begin{equation}
  M = \sum_{i=1}^{N}S_{i},
\end{equation}
$<\cdots >_{{T},\{S\}}$ denotes the thermal average
for the $"S"$ spins at temperature, $T$,
and $Z(K,\{\tau \})$ is the partition function 
at the inverse temperature, $K(=1/T)$, with the bond configuration, 
$\{\tau \}$ . We show the summary of the 
derivation of equation (6) in the
appendix. (For the details of the derivation,
see [19].)
When we define the modified probability
distribution, $P_{2}(K,K_{p},\{\tau \})$,
 for the bond configuration,
$\{\tau \}$, as
\begin{equation}
 P_{2}(K,K_{p},\{\tau \}) = \frac {1}{(2\cosh (K_{p}))^{N_{\rm B}}}
   \exp (K\sum_{(ij)}\tau_{ij})
    \frac {Z(K_{p},\{\tau \})}{Z(K,\{\tau \})},
\end{equation}
we can then write
\begin{equation}
   [<Q^{2n}>_{{T},\{S^{\alpha},S^{\beta}\}}]_{p}=
    \{<M^{2n}>_{{T},\{S\}}\}_{K,K_{p}},
\end{equation}
where $\{\cdots \}_{K,K_{p}}$ denotes the configurational average by the
modified bond probability  distribution.
That is, $[<Q^{2n}>_{{T},\{S^{\alpha},S^{\beta}\}}]_{p}$ at 
temperature, $T$, with
the ferromagnetic bond concentration, $p$, is transformed 
into the configurational average
of $<M^{2n}>_{{T},\{S\}}$ by  the modified bond  probability disribution,
$P_{2}(K,K_{p},\{\tau \})$.
Similarly, we can get the following identity \cite{Kitatani}:
\begin{equation}
   [<M^{2n}>_{{T},\{S\}}]_{p}=\{<M^{2n}>_{{T_{p},\{S\}}}\}_{K,K_{p}},
\end{equation}
where $T_{p}=1/K_{p}$ .
(Note that the above argument applies to
any dimension.)

  Hereafter, we explain  a new approach  to numerically calculate
the values of $[<Q^{2n}>_{{T},\{S^{\alpha},S^{\beta}\}}]_{p}$,
 using equation (9).
To realize the bond configuration with the modified bond  probability
distribution, $P_{2}(K,K_{p},\{\tau \})$,
 we use the conventional Monte Carlo technique.
We define $W(\tau_{ij}\rightarrow -\tau_{ij},\{\tau \}')$
as the transition probability by which the value of the bond, 
$\tau_{ij}$, changes.
 To guarantee that the stationary probability distribution  becomes
$P_{2}(K,K_{p},\{\tau_{ij}\})$, the following detailed
 balance must be satisfied:
\begin{eqnarray}
\fl P_{2}(K,K_{p},\tau_{ij},\{\tau \}')W(\tau_{ij}
\rightarrow -\tau_{ij},\{\tau \}')  \nonumber\\
=P_{2}(K,K_{p},-\tau_{ij},\{\tau \}')W(-\tau_{ij}
\rightarrow \tau_{ij},\{\tau \}')
\end{eqnarray}
 Using equation (8), we obtain
\begin{equation}
\fl \frac {W(\tau_{ij}\rightarrow -\tau_{ij},\{\tau \}')}
       {W(-\tau_{ij}\rightarrow \tau_{ij},\{\tau \}')} 
   = \exp (-2K\tau_{ij}) 
    \frac { \cosh (2K_{p})-\sinh (2K_{p})<\tau_{ij}S_{i}S_{j}>_{
          {T_{p}},\{S\}}}
           { \cosh (2K) - \sinh (2K)
           <\tau_{ij}S_{i}S_{j}>_{{T},\{S\}}}
\end{equation}
Namely, when we can calculate 
$<S_{i}S_{j}>_{{T},\{S\}}$  in a particular
bond configuration, we can estimate the transition probability.

The processes to calculate
 $[<Q^{2n}>_{{T},\{S^{\alpha},S^{\beta}\}}]_{p}$
are as follows:\\
1)We start from an arbitrary bond configuration.\\
2)Using the transfer matrix method, we exactly
 calculate the value of $<S_{i}S_{j}>_{{T},\{S\}}$,
 and successively flip the bond, $\tau_{ij}$,
 according to the transition probability,
$W(\tau_{ij}\rightarrow -\tau_{ij},\{\tau \}')$, using
the conventional Monte Carlo technique.\\
3)We continue  process 2)
until the modified bond  probability distribution, $P_{2}(K,K_{p},\{\tau \})$
is realized.\\
4)We calculate the value of $<M^{2n}>_{{T},\{S\}}$ for
 each bond configuration
using the transfer matrix method.\\
5)We repeat  processes, 2) and 4).\\
6)Finally, the simple average of $<M^{2n}>_{{T},\{S\}}$ 
gives the value of 
$[<Q^{2n}>_{{T},\{S^{\alpha},S^{\beta}\}}]_{p}$ with 
the bond probability distribution,
 $P(\tau_{ij})$.\\

We now show the efficiency of this method.
We define $n_{a}$, $n_{b}$ and $n_{c}$
as the number of  initial Monte Carlo skip steps, 
the number of  Monte Carlo steps where we calculate
$<M^{2n}>_{{T},\{S\}}$, and the number of independent Monte Carlo runs,
respectively. 
In all the calculations, we have evaluated  
$\{\tau_{ij}(0)\tau_{ij}(t)\}_{K,K_{p}}$ using the statistical
 dependence time method \cite{Ito}, and find that
the relaxation time of 
 $\{\tau_{ij}(0)\tau_{ij}(t)\}_{K,K_{p}}$
is always very small even when compared with one Monte
Carlo step time. 
That is, only several tens of initial skip steps are enough to
realize the stationary bond probability distribution.
For example, we have compared the values of
the spin-glass susceptibility $\chi_{\rm SG}
(=[<Q^{2}>_{{T},\{S^{\alpha},S^{\beta}\}}]_{p}/N)$ 
calculated by the present method and that by the 
conventional transfer matrix method using real replicas.
Table 1 shows the results at $T=0.1$ for $L=7$. 
The calculations by the
conventional transfer matrix method have been done
 with $10^{5}$ independent bond
configurations. The error bars of the present methods
have been estimated in the same way as those of
 conventional Monte Carlo simulations.
From table 1, we can see that all the data are consistent
within the error bars, and the size of the error bars of all
the calculations are of the same magnitude.
Consequently, we find that only 20 steps are enough
for the initial Monte Carlo skip steps. 
Furthermore, we have examined whether equation (9) holds
 or not at $p=0.5-0.95$, $T=0.1-0.5$
for $L=7$. We have also examined whether equation (10) holds or 
not at $p=0.8-0.9$,
$T=0.1-0.4$ for $L=15$.
All the results are consistent in a statistical sense, from which
we conclude  that the present method is usable and not affected by
systematic errors.

\begin{table}
\caption{The values of $\chi_{\rm SG}$ with
various ($n_{a},n_{b},n_{c}$) at $T=0.1$ for $L=7$.
The value of $\ast$ is calculated by the conventional
transfer matrix method using real replicas with $10^{5}$
bond samples.}
\begin{indented}
\item[]\begin{tabular}{@{}ll}
\br
($n_{a},n_{b},n_{c}$)&$\chi_{\rm SG}$\\
\mr
(2000,10000,10)&29.076(24)\\
(200,1000,100)&29.084(38)\\
(20,100,1000)&29.094(35)\\
(20,20,5000)&29.045(34)\\
$\ast$&29.037(35)\\
\br
\end{tabular}
\end{indented}
\end{table}

\section{The spin-glass phase transition of the two-dimensional
$\pm J$ Ising model}
We have extensively investigated
 the two-dimensional $\pm J$ Ising model 
 for $p=0.5$.
The  results of the asymmetric case $(p>0.5)$
will be given in a subsequent paper \cite{Kitatani2}.
For $p=0.5$, we have calculated 
$[<Q^{2n}>_{{T},\{S^{\alpha},S^{\beta}\}}]_{p}$ at $T=0.1-0.5$
with the linear size $L=7-23$.
The calculations have been performed with 
$(n_{a},n_{b},n_{c})=(200,1000,100)$  for $L\leq 21$
and (200,200,480) for $L=23$.

The energy gap between the ground state and the first
excitation state is two in the unit of the
interaction strength. Thus, in finite systems, 
any physical quantity at a very low temperature must saturate to
its value at $T=0$.
We show  the temperature
dependence of the spin-glass susceptibility, $\chi_{\rm SG}$,
in figure 1. We find, indeed, that
the values of $\chi_{\rm SG}$ for each $L$ show the strong saturation near 
$T=0$, and the tendency becomes clearer as the system size becomes smaller,
as has already been  pointed out by several authors
\cite{Bhatt2,Shirakura,Shirakura2}.
The Binder parameter \cite{Binder}
\begin{equation}
       g_{L}=\frac {1}{2}
     (3-\frac {[<Q^{4}>_{{T},\{S^{\alpha},S^{\beta}\}}]_{p}}
{[<Q^{2}>_{{T},\{S^{\alpha},S^{\beta}\}}]^{2}_{p}})
\end{equation}
is widely used for the determination of the critical temperature.
The value of the Binder parameter becomes asymptotically
size independent  for
large systems  at the critical temperature. Therefore,
the point where this quantity becomes asymptotically size independent 
gives an estimation of the critical point.
The simple plot of the Binder parameter versus temperature
is shown in figure 2. 
We can clearly see that  the data for different sizes intersect 
at almost the same temperature, $T=0.3$, and the size dependence
of the intersection points is very small. We cannot, however, 
immediately conclude
that the spin-glass phase transition occurs at $T\simeq 0.3$,
 since the intersection might be due to 
 the strong saturation of the data
near $T=0$ \cite{Bhatt2,Shirakura,Shirakura2}.
Therefore, we perform  scaling analyses for $g_{L}$ and $\chi_{\rm SG}$.
There is no general rule
to avoid the disturbance from the saturation mentioned above
in the scaling analyses. Here, we adopt a criterion that
every data point is not used all through the scaling analyses,
when the value of $\chi_{\rm SG}$ increases less than $3\%$ in the temperature
interval, $\Delta T=0.05$. 
Although the  criterion, which
we have determined from the observation in figure 1, seems rather artificial, 
we believe
that this criterion systematically removes the saturation to $T=0$ 
in a certain sense.
 Consequently, we use, for example,
the data with $T\ge 0.35$ for $L=7$, and with $T\ge 0.2$ for $L=23$.

First, we perform the scaling analyses without including
the corrections to scaling.
In this case,
the Binder parameter, $g_{L}$, has the scaling form
\begin{equation}
      g_{L} = \bar{g}(L^{1/\nu }(T-T_{\text{c}})),
\end{equation}
and that of the spin-glass susceptibility, $\chi_{\rm SG}$, is
\begin{equation}
     \chi_{\rm SG} = L^{2-\eta}\bar{\chi}(L^{1/\nu }(T-T_{\text{c}})),
\end{equation}
where $\nu$ is the critical exponent of the spin-glass correlation 
length, and
$\eta$ is the critical exponent which describes the decay of the 
correlation at the critical temperature.
 Figure 3 shows the best-fit scaling plot of the
Binder parameter, which indicates that $T_{\text{c}}\simeq 0.3$
and the critical exponent $\nu \simeq 1.3$.
We can see that  the  data at $T<T_{\text{c}}=0.3$
and $T\geq T_{\text{c}}$ fit
rather well on one scaling function,
which
indicates  that the spin-glass phase
transition of the two-dimensional $\pm J$ Ising model
is a conventional phase transition, and there exists
a finite long range order at $T<T_{\text{c}}$.
The best-fit scaling plot of the spin-glass susceptibility  is
also  shown in figure 4, which
indicates  that the critical exponent $\eta \simeq 0.225$. 
The estimated values of $T_{\rm c}$ and the critical exponents
are similar to those determined 
by Shirakura {\it et al}\cite{Shirakura}.
Figures 5 and 6 show the
scaling plots of the Binder parameter
and the spin-glass susceptibility, assuming $T_{\rm c}=0$, $\nu =2.6$
and $\eta =0.2$ \cite{Bhatt2},
where we clearly see  the systematic deviations.
 Namely, the scaling plots without
including the  corrections to scaling
strongly indicate the existence of 
a finite-temperature spin-glass phase transition. 

Next, we perform the scaling analyses including the corrections to scaling.
We take the scaling forms of the Binder paremeter and
the spin-glass susceptibility as
\begin{equation}
         g_{L}=\bar{g}(L^{1/\nu }(T-T_{\text{c}}))(1+\frac {a}
         {L^{\omega}}),
\end{equation}
and 
\begin{equation}
     \chi_{\rm SG} = L^{2-\eta}\bar{\chi}(L^{1/\nu }(T-T_{\text{c}}))
     (1+\frac {b}{L^{\omega '}}).
\end{equation}
There is little quantitative change in figures 3 and 4, even
though we fit the data using equations (16) and (17). Thus, when
we assume $T_{\rm c}=0.3$, 
the effect of the
corrections to scaling is rather small. 
On the other hand, assuming that 
$T_{\rm c}=0$, $\nu =2.6$ and $\eta =0.17$ \cite{Bhatt2},
the data of the Binder parameter and the
spin-glass susceptibility  fit very well on
one scaling function, respectively, as shown in figures 7 and 8
with $\omega =\omega '=0.5$ and $a=b=-0.3$, although
the data with a small linear size, $L=7$, deviate from the
scaling function. (To fit the data, we use $\eta =0.17$,
which is, for example, consistent with the result in [8], $\eta=
0.2\pm 0.05$.)
Thus, including the corrections to scaling, both $T_{\rm c}=0$
and $T_{\rm c} \simeq 0.3$ are consistent with the scaling analyses.
Furthermore, we find that every temperature for $0\leq T\le 0.3$  
might  become the critical temperature, $T_{\rm c}$,
in this scaling form. 
Consequently,
we find that the estimation of the value of $T_{\rm c}$
is strongly affected by the corrections to scaling
in the two-dimensional $\pm J$ Ising model
within
the present data ($L\leq 23$). 

\section{Conclusions}
We have developed a new method to numerically calculate 
$[<Q^{2n}>_{{T},\{S^{\alpha},S^{\beta}\}}]_{p}$ of the 
two-dimensional $\pm J$ Ising model,
where, using a local gauge transformation,
 $[<Q^{2n}>_{{T},\{S^{\alpha},S^{\beta}\}}]_{p}$ 
can be calculated as a simple average of
$<M^{2n}>_{{T},\{S\}}$ with a modified bond probability distribution.
By using the  present method, we have extensively
calculated the values of
 $[<Q^{2n}>_{{T},\{S^{\alpha},S^{\beta}\}}]_{p}$,
where the statistical errors become about
 an order of magnitude  smaller than in
previous studies, and we have investigated the scaling analyses
including the corrections to scaling.
 By using the scaling analyses without
including the corrections to scaling,
our data strongly indicate a finite-temperature
spin-glass phase transition.
We find, however, that the estimation of $T_{\text c}$ is strongly
affected by the corrections to scaling within
the data with $L\leq 23$,
and that every temperature for $0\leq T\le 0.3$
might be able to become the critical temperature.
Consequently, our results indicate that there still
remains the possibility that $T_{\text c}=0$,
contrary to the recent results of Shirakura {\it et al}
\cite{Shirakura,Shiomi,Matsubara} which suggest the existence of 
a finite-temperature spin-glass phase transition.

\ack
The authors  thank Dr. Shirakura for his 
useful comments and discussions.
The calculations were made on the FACOM VPP500 at the Institue for
Solid State Physics, University of Tokyo and
on the HITAC SR8000 at the University of Tokyo.
This work was partially supported by a Grant-in-Aid
 for Exploratory Research from
the Ministry of Education, Science, Sports and Culture.

\begin{appendix}
\section{}

In this appendix,
we briefly explain the derivation of equation (6).
We show that both sides of equation (6) coincide with
each other.

Using equation (5), $[<Q^{2n}>_{{T},\{S^{\alpha},S^{\beta}\}}]_{p}$
is written as
\begin{equation}
\fl  [<Q^{2n}>_{{T},\{S^{\alpha},S^{\beta}\}}]_{p}=\frac {1}{C}
   \sum_{\tau_{ij}=\pm 1}\exp (K_{p}\sum_{(ij)}\tau_{ij})
 <(\sum_{i=1}^{N} S_{i}^{\alpha }S_{i}^{\beta })^{2n}>
  _{{T},\{S^{\alpha},S^{\beta}\}},
\end{equation}
and we abbreviate $(2\cosh (K_{p}))^{N_{B}}$ as $C$ from now on. 

Here, we perform the
following local gauge transformation:
\begin{equation}
  \tau_{ij} \rightarrow \tau_{ij}\sigma_{i}\sigma_{j},\;
  S_{i}^{\alpha } \rightarrow S_{i}^{\alpha }\sigma_{i},\; 
  S_{i}^{\beta } \rightarrow S_{i}^{\beta }\sigma_{i},\;                                                       (\sigma_{i}=\pm 1)
\end{equation}
where each $\sigma_{i}$ arbitrarily takes $+1$ or $-1$.
Since
$<(\sum_{i=1}^{N} S_{i}^{\alpha }S_{i}^{\beta })^{2n}>
_{{T},\{S^{\alpha},S^{\beta}\}}$
is
invariant under this transformation,
we obtain
\begin{equation}
\fl  [<Q^{2n}>_{{T},\{S^{\alpha},S^{\beta}\}}]_{p}=\frac {1}{C}
   \sum_{\tau_{ij}=\pm 1}\exp (K_{p}\sum_{(ij)}\tau_{ij}
   \sigma_{i}\sigma_{j})
   <(\sum_{i=1}^{N} S_{i}^{\alpha }S_{i}^{\beta })^{2n}>
    _{{T},\{S^{\alpha},S^{\beta}\}},
\end{equation}
where we note that the summation over $\tau_{ij}
\sigma_{i}\sigma_{j}=\pm 1$
is eqivalent to  that over $\tau_{ij}=\pm 1$.
As each $\sigma_{i}$ arbitarily takes $+1$ or $-1$,
therefore,
we take all the summations of ${\sigma_{i}}$ and
divide by $2^{N}$,
namely
\begin{eqnarray}
\fl  [<Q^{2n}>_{{T},\{S^{\alpha},S^{\beta}\}}]_{p}=\frac {1}{C}
   \frac{1}{2^{N}} \sum_{\sigma_{i}=\pm 1}
    \sum_{\tau_{ij}=\pm 1}\exp (K_{p}\sum_{(ij)}\tau_{ij}
  \sigma_{i}\sigma_{j})
 <(\sum_{i=1}^{N} S_{i}^{\alpha }S_{i}^{\beta })^{2n}>
  _{{T},\{S^{\alpha},S^{\beta}\}} \nonumber\\
\lo =\frac {1}{C} \sum_{\tau_{ij}=\pm 1}
  \frac{Z(K_{p},\{\tau \} )}{2^{N}}
  <(\sum_{i=1}^{N} S_{i}^{\alpha }S_{i}^{\beta })^{2n}>
   _{{T},\{S^{\alpha},S^{\beta}\}}
\end{eqnarray}

Next, we consider the r.h.s. (we denote it as A) of equation (6).
The r.h.s. of equation (6) is written as
\begin{eqnarray}
\fl A=\frac {1}{C}
   \sum_{\tau_{ij}=\pm 1}\exp (K\sum_{(ij)}\tau_{ij})
    \frac {Z(K_{p},\{\tau \})}{Z(K,\{\tau \})}
    <(\sum_{i=1}^{N}S_{i})^{2n}>_{{T},\{S\}}  \nonumber\\
\fl    =\frac {1}{C}
   \sum_{\tau_{ij}=\pm 1}\exp (K\sum_{(ij)}\tau_{ij})
    \frac {Z(K_{p},\{\tau \})}{Z(K,\{\tau \})}
   \frac{ \sum_{S_{i}=\pm 1}\exp (K\sum_{(ij)}\tau_{ij}S_{i}S_{j})
    (\sum_{i=1}^{N}S_{i})^{2n}}{Z(K,\{\tau \})}
\end{eqnarray}

Here, we perform the same local gauge transformation.
Then, we obtain
\begin{eqnarray}
\fl A = \frac {1}{C}
   \sum_{\tau_{ij}=\pm 1}\exp (K\sum_{(ij)}\tau_{ij}
   \sigma_{i}\sigma_{j})
    \frac {Z(K_{p},\{\tau \})}{Z(K,\{\tau \})}
 \frac{ \sum_{S_{i}=\pm 1}\exp (K\sum_{(ij)}\tau_{ij}S_{i}S_{j})
    (\sum_{i=1}^{N}S_{i}\sigma_{i})^{2n}}{Z(K,\{\tau \})} \nonumber\\
\fl = \frac {1}{C}
   \sum_{\tau_{ij}=\pm 1}Z(K_{p},\{\tau \})
 \frac {\exp (K\sum_{(ij)}\tau_{ij}\sigma_{i}\sigma_{j})
     \sum_{S_{i}=\pm 1}\exp (K\sum_{(ij)}\tau_{ij}S_{i}S_{j})
     (\sum_{i=1}^{N}S_{i}\sigma_{i})^{2n}}
     {Z(K,\{\tau \})^{2}} \nonumber\\
\fl   =\frac {1}{C}
     \sum_{\tau_{ij}=\pm 1}\frac {Z(K_{p},\{\tau \})}{2^{N}}
 \frac {\sum_{S_{i},\sigma_{i}=\pm 1}
     \exp (K\sum_{(ij)}\tau_{ij}\sigma_{i}\sigma_{j})
     \exp (K\sum_{(ij)}\tau_{ij}S_{i}S_{j})
     (\sum_{i=1}^{N}S_{i}\sigma_{i})^{2n}}
     {Z(K,\{\tau \})^{2}}  \nonumber\\
\fl    =\frac {1}{C}
     \sum_{\tau_{ij}=\pm 1}\frac {Z(K_{p},\{\tau \})}{2^{N}}
     <(\sum_{i=1}^{N}S_{i}\sigma_{i})^{2n}>
     _{{T},\{S,\sigma \}}.
\end{eqnarray}
 Thus, from equations (A4) and (A6), we conclude that
 both sides of equation (6) 
coincide with each other.

\end{appendix}

\clearpage

\twocolumn

\begin{figure}[p]
\epsfxsize=60mm
\epsfbox{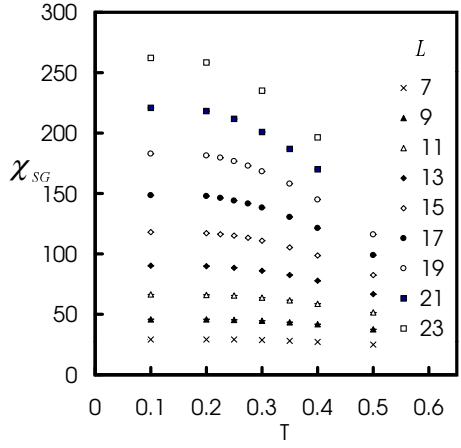}
\caption{A plot of $\chi_{\rm SG}$ versus $T$.}
\end{figure}

\begin{figure}[p]
\epsfxsize=60mm
\epsfbox{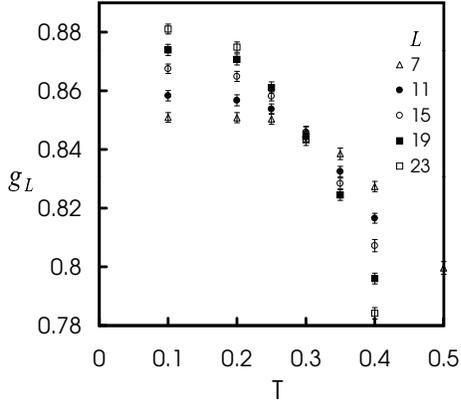}
\caption{A plot of $g_{L}$ versus $T$.}
\end{figure}

\begin{figure}[p]
\epsfxsize=60mm
\epsfbox{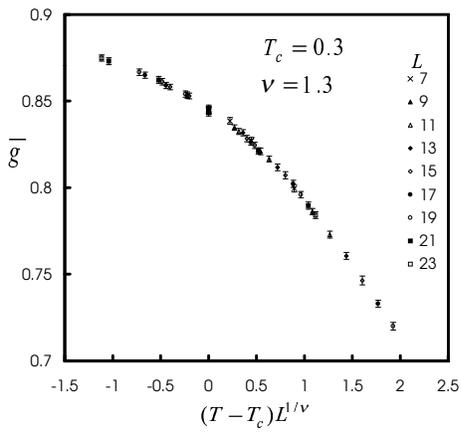}
\caption{A scaling plot for $g_{L}$.}
\end{figure}

\begin{figure}[p]
\epsfxsize=60mm
\epsfbox{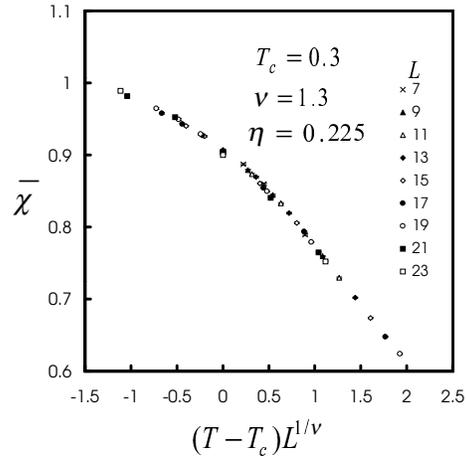}
\caption{A scaling plot for $\chi_{\rm SG}$.}
\end{figure}

\begin{figure}[p]
\epsfxsize=60mm
\epsfbox{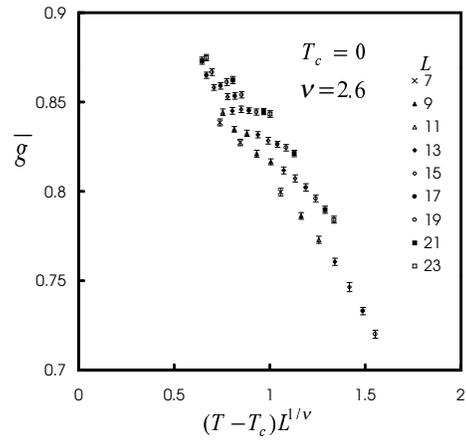}
\caption{A scaling plot for $g_{L}$, assuming $T_{\rm c}=0$.}
\end{figure}

\begin{figure}[p]
\epsfxsize=60mm
\epsfbox{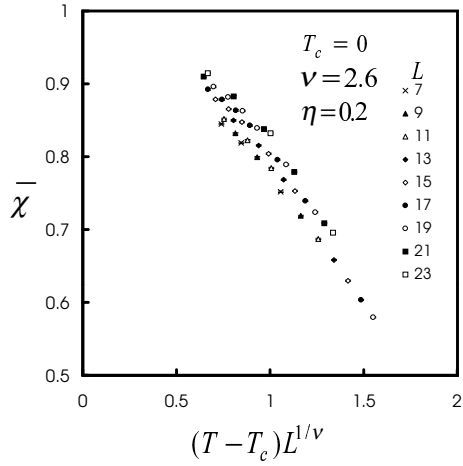}
\caption{A scaling plot for $\chi_{\rm SG}$, assuming $T_{\rm c}=0$.}
\end{figure}

\begin{figure}[p]
\epsfxsize=60mm
\epsfbox{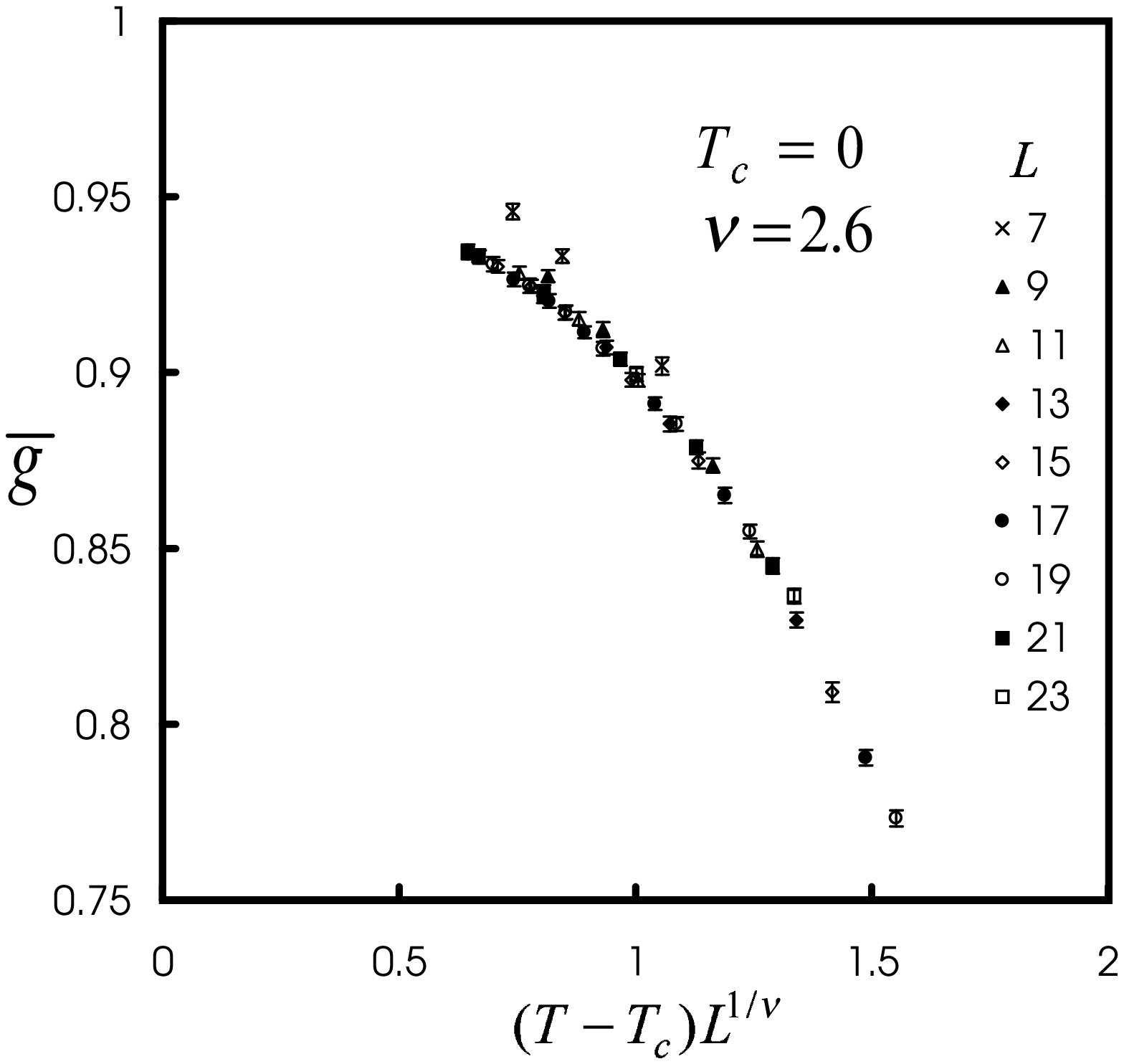}
\caption{A scaling plot for $g_{L}$ including the corrections to
scaling with $\omega =0.5$ and $a=-0.3$, assuming $T_{\rm c}=0$.}
\end{figure}

\begin{figure}[p]
\epsfxsize=60mm
\epsfbox{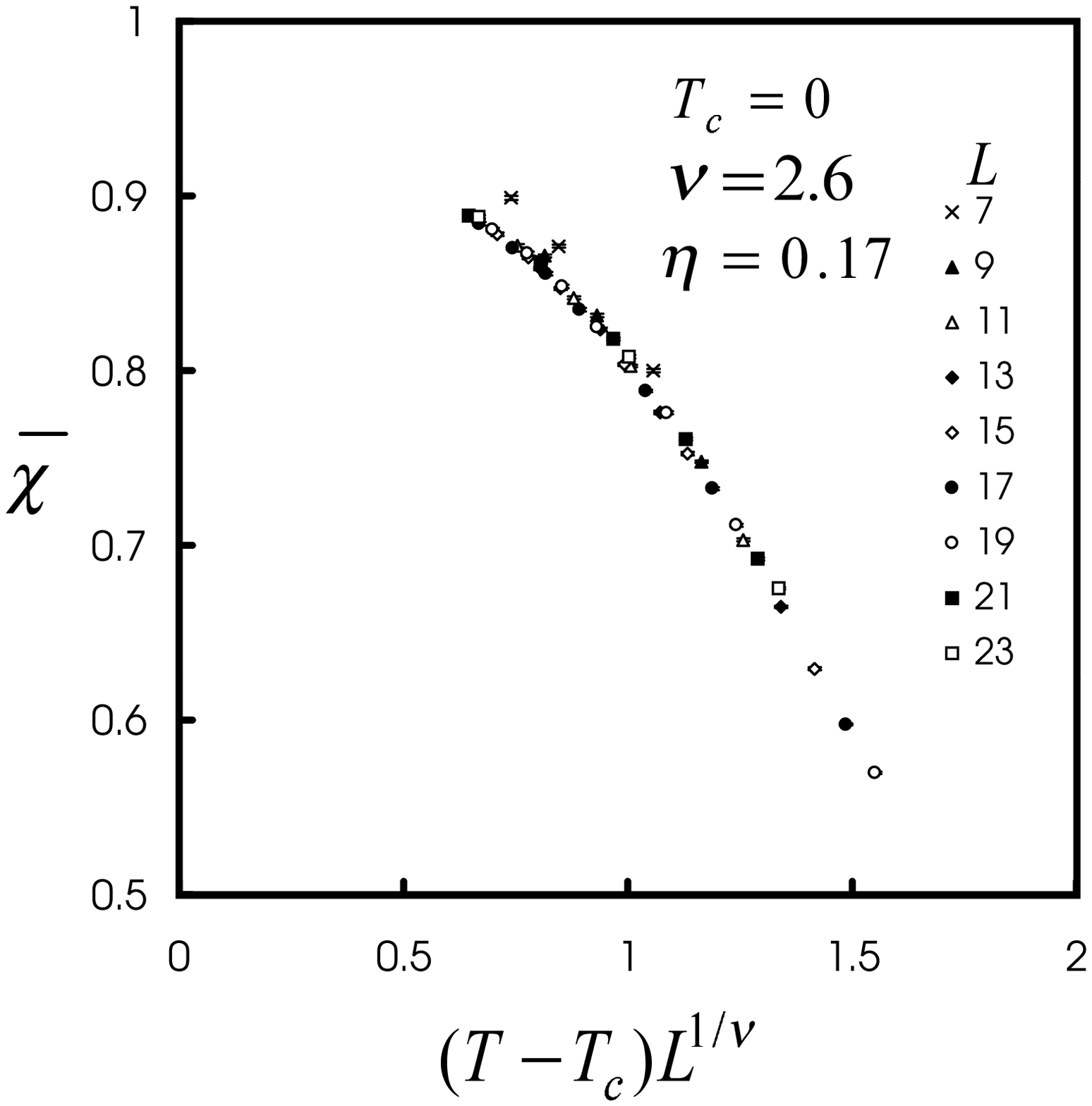}
\caption{A scaling plot for $\chi_{\rm SG}$ including the corrections to
scaling with $\omega ' =0.5$ and $b=-0.3$, assuming $T_{\rm c}=0$.}
\end{figure}


\section*{References}
\begin{thebibliography}{100}
\bibitem{Bhatt} Bhatt R N and Young A P 1985
{\it Phys. Rev. Lett.} {\bf 54} 924
\bibitem{Ogielski} Ogielski A T and Morgenstern I 1985
{\it Phys. Rev. Lett.} {\bf 54} 928
\bibitem{Ogielski2} Ogielski A T 1985
{\it Phys. Rev.} B {\bf 32} 7384 
\bibitem{Kawashima} Kawashima N and Young A P 1996
{\it Phys. Rev.} B {\bf 53} R484 
\bibitem{Singh} Singh R R P and  Chakravarty S 1986
{\it Phys. Rev. Lett.} {\bf 57} 245 
\bibitem{Bray} Bray A J and Moore M A 1984
{\it J. Phys.} C {\bf 17} L463
\bibitem{Swendsen} Swendsen R H and Wang J -S 1986
{\it Phys. Rev. Lett.} {\bf 57} 2607 
\bibitem{Bhatt2} Bhatt R N and Young A P 1988
{\it Phys. Rev.} B {\bf 37} 5606
\bibitem{Morgenstern} Morgenstern I and Binder K 1980
{\it Phys. Rev.} B {\bf 22} 288 
\bibitem{Ozeki2} Ozeki Y 1990
{\it J. Phys. Soc. Jpn.} {\bf 59} 3531
\bibitem{Kawashima2} Kawashima N and Rieger H 1997
{\it Europhys. Lett.} {\bf 39} 85
\bibitem{Shirakura} Shirakura T and Matsubara F 1996
{\it J. Phys. Soc. Jpn.} {\bf 65} 3138
\bibitem{Shiomi} Shiomi M, Matsubara F and Shirakura T
{\it private communication}
\bibitem{Shirakura2} Shirakura T and Matsubara F 1997
{\it Phys. Rev. Lett.} {\bf 79} 2887
\bibitem{Matsubara} Matsubara F, Shirakura T
and Shiomi M 1998 {\it Phys. Rev.} B {\bf 58} R11821
\bibitem{Ozeki} Ozeki Y and Nishimori H 1987
{\it J. Phys. Soc. Jpn.} {\bf 56} 3265
\bibitem{Kitatani3} Kitatani H and Oguchi T 1992
{\it J. Phys. Soc. Jpn.} {\bf 61} 1598
\bibitem{Nishimori} Nishimori H 1981
{\it Prog. Theor. Phys.} {\bf 66} 1169
\bibitem{Kitatani} Kitatani H 1992
{\it J. Phys. Soc. Jpn.} {\bf 61} 4049
\bibitem {Kitatani2} Kitatani H and Seike T
{\it in preparation}
\bibitem {Binder} Binder K 1991
{\it Phys. Rev. Lett.} {\bf 47} 119
\bibitem{Ito} Kikuchi M and Ito N 1993
{\it J. Phys. Soc.Jpn.} {\bf 62} 3052 
\end{thebibliography}
\end{document}